\documentclass[conference]{IEEEtran}
\ifCLASSINFOpdf
\else
\fi

\usepackage{graphicx} 
\usepackage{amsmath,amsfonts,amsthm}
\usepackage[]{mdframed}

\usepackage{amsmath,amsfonts,bm}

\def\1{\bm{1}}

\def\vk{{\bm{k}}}

\def\vv{{\bm{v}}}
\def\vw{{\bm{w}}}
\def\vx{{\bm{x}}}
\def\vy{{\bm{y}}}

\def\mI{{\bm{I}}}

\DeclareMathAlphabet{\mathsfit}{\encodingdefault}{\sfdefault}{m}{sl}
\SetMathAlphabet{\mathsfit}{bold}{\encodingdefault}{\sfdefault}{bx}{n}

\def\gF{{\mathcal{F}}}

\def\gK{{\mathcal{K}}}

\def\gM{{\mathcal{M}}}
\def\gN{{\mathcal{N}}}

\def\gU{{\mathcal{U}}}

\newcommand{\R}{\mathbb{R}}

\newcommand{\T}{\top}

\newtheorem{myLem}{Lemma}
\newtheorem{myThm}{Theorem}

\usepackage{cleveref}
\Crefname{myLem}{Lemma}{Lemmas}
\usepackage{xcolor}

\begin{document}
%
\title{Towards a Sampling Theory for Implicit Neural Representations}

\author{\IEEEauthorblockN{Mahrokh Najaf\qquad Gregory Ongie}
\IEEEauthorblockA{Department of Mathematical and Statistical Sciences\\
Marquette University\\
Milwaukee, WI}}


%


\maketitle

\begin{abstract}
Implicit neural representations (INRs) have emerged as a powerful tool for solving inverse problems in computer vision and computational imaging. INRs represent images as continuous domain functions realized by a neural network taking spatial coordinates as inputs. However, unlike traditional pixel representations, little is known about the sample complexity of estimating images using INRs in the context of linear inverse problems. Towards this end, we study the sampling requirements for recovery of a continuous domain image from its low-pass Fourier coefficients by fitting a single hidden-layer INR with ReLU activation and a Fourier features layer using a generalized form of weight decay regularization. Our key insight is to relate minimizers of this non-convex parameter space optimization problem to minimizers of a convex penalty defined over a space of measures. We identify a sufficient number of samples for which an image realized by a width-1 INR is exactly recoverable by solving the INR training problem, and give a conjecture for the general width-$W$ case.  To validate our theory, we empirically assess the probability of achieving exact recovery of images realized by low-width single hidden-layer INRs, and illustrate the performance of INR on super-resolution recovery of more realistic continuous domain phantom images.
\end{abstract}

\IEEEpeerreviewmaketitle

\section{Introduction}
Many inverse problems in signal and image processing are naturally posed in continuous domain. For example, in magnetic resonance imaging, measurements are modeled as samples of the Fourier transform of a function $f(\vx)$ defined over continuous spatial coordinates $\vx \in \R^3$ \cite{fessler2010model}. Often these continuous domain inverse problems are approximated using discrete image or signal representations, along with an appropriate discretization of the forward model, such as the discrete Fourier transform in place of the continuous Fourier transform. However, this can lead to undesirable discretization artifacts, such as Gibb's ringing \cite{gelb2002,veraart2016gibbs,ongie2015,ongie2016off}. Such artifacts may be reduced by increasing the resolution of the discretization grid, but at the cost of additional computational complexity and potentially making the problem ill-posed without additional regularization.

Recently, continuous domain image representations using neural networks have gained traction in a variety of inverse problems arising in computer vision and computational imaging (see, e.g., the recent review \cite{xie2022neural}). These so-called \emph{implicit neural representations} (INRs), also variously known as 
\emph{coordinate-based neural networks} or \emph{neural radiance fields}, parameterize a signal or image as a neural network defined over continuous spatial coordinates \cite{mildenhall2021nerf,sitzmann2020implicit,tancik2020fourier}. The main benefits of INRs over discrete representations are that (1) they offer a compact (non-linear) representation of signals/images with far fewer parameters than a discrete representation, (2) they allow for more accurate continuous domain forward modeling, which can reduce the impact of discretization artifacts, and (3) standard neural network training can be used to fit an INR to measurements, sidestepping the need for custom iterative solvers.

However, the fundamental limits of image/signal estimation using INRs have not been explored in detail. One basic unanswered question is: How many linear measurements are necessary and sufficient for accurate reconstruction of a signal/image by fitting an INR? Additionally, how does the architecture of the INR influence the recovered signal/image?  As a step in the direction of answering these questions, we consider the recovery of a signal/image from its low-pass Fourier coefficients using INRs. In particular, we focus on a popular INR architecture originally introduced in \cite{tancik2020fourier} that combines a fully-connected neural network with a Fourier features layer.

Recent work has shown that training a single hidden-layer ReLU network with $\ell^2$-regularization on its parameters (also known as ``weight decay'' regularization) imparts a sparsity effect, such that the network learned has low effective width \cite{savarese2019infinite,ongie2019function,parhi2021banach,parhi2023deep}. Bringing this perspective to INRs, we study the conditions under which a signal/image realizable by a single-hidden layer, low-width INR is exactly recoverable from low-pass Fourier samples by minimizing a generalized weight decay objective. Our main theoretical insight is to relate minimizers of this non-convex parameter space optimization problem to the minimizers of a convex optimization problem posed over a space of measures. Using this framework, we identify a sufficient number of samples such that an image realizable by a width-1 INR is the unique minimizer of the INR training problem, and state a conjecture for widths greater than one. We validate this conjecture with numerical simulations. Finally, we illustrate the practical performance of INRs on super-resolution recovery of continuous domain phantom images, highlighting the impact of parameter-space regularization and network depth on recovery performance.

\section{Problem formulation}
We consider the recovery of a $d$-dimensional real-valued continuous domain signal/image $f:[0,1]^d\rightarrow \R$ from a sampling of its Fourier coefficients $\hat f$ defined by
\[
\hat{f}[\vk] = \int_{[0,1]^d}f(\vx)e^{-2\pi i \vk^\top \vx}d\vx,
\]
where $\vk\in \mathbb{Z}^d$ is any $d$-tuple of integer spatial frequencies. In particular, suppose we have access to all low-pass Fourier coefficients belonging to a uniformly sampled square region in frequency domain
$
\Omega = \{\vk \in \mathbb{Z}^d : \|\vk\|_\infty \leq K \}
$
where $K$ is a positive integer.
Let $\gF_\Omega f = (\hat{f}[\vk])_{\vk \in \Omega}$ denote the vector of Fourier coefficients of $f$ restricted to $\Omega$. We study the inverse problem of recovering the function $f$ given $\vy = \gF_\Omega f$. 

Clearly, this is an ill-posed inverse problem if no further assumptions are made on $f$. The INR approach attempts to resolve this ill-posedness by assuming $f$ is well-approximated by a function realized by a ``simple'' neural network $f_\theta$ with trainable parameters $\theta$. Under this assumption, one can attempt to recover $f$ from measurements $\vy$ by solving the (non-linear) least squares problem
\begin{equation}\label{eq:lsfit_noreg}
    \min_{\theta} \tfrac{1}{2}\|\gF_\Omega f_\theta - \vy\|_2^2.
\end{equation}
However, this problem may still be ill-posed (i.e., \eqref{eq:lsfit_noreg} may have multiple solutions) unless the number of trainable parameters used to define the INR $f_\theta$ is heavily constrained. Yet, useful solutions of \eqref{eq:lsfit_noreg} may be found by relying on implicit regularization induced by practical gradient-based algorithms \cite{gunasekar2018characterizing}.

As an alternative to using an overly constrained network architecture or relying on implicit regularization to resolve ill-posedness, we propose incorporating an explicit parameter space regularizer $R(\theta)$ to the least squares objective:
\begin{equation}\label{eq:lsfit}
    \min_{\theta} \tfrac{1}{2}\|\gF_\Omega f_\theta - \vy\|_2^2 + \lambda R(\theta),
\end{equation}
where $\lambda > 0$ is a regularization parameter.
In particular, we focus on a class of regularizers $R(\theta)$ that generalize weight decay regularization, i.e., the squared $\ell^2$-norm of all trainable parameters; this class is described in more detail below in Section \ref{sec:weight_decay}. 

To tightly enforce data consistency constraints, the regularization parameter $\lambda$ in \eqref{eq:lsfit} should be very small.  As a model for this situation, we will focus on the equality constrained problem
\begin{equation}\label{eq:opt_param_space}
\min_{\theta} R(\theta)~~s.t.~~\gF_\Omega f_\theta = \vy,
\end{equation}
which can be thought of as the limiting case of \eqref{eq:lsfit} as $\lambda\rightarrow 0$. Our main goal is to characterize global minimizers of  \eqref{eq:opt_param_space}. In particular, we are interested in its \emph{function space minimizers}, i.e., functions $f_{\theta^*}$ where $\theta^*$ is a global minimizer of \eqref{eq:opt_param_space}. In particular, given $\vy = \gF_\Omega f$ where $f$ is realizable as an INR, we ask: when is $f$ the unique function space minimizer of \eqref{eq:opt_param_space}? Next, we describe the architectural assumptions we put on the INR $f_\theta$ and the associated regularizer $R(\theta)$.

\begin{figure}[ht!]
    \centering
    \includegraphics[width=\columnwidth]{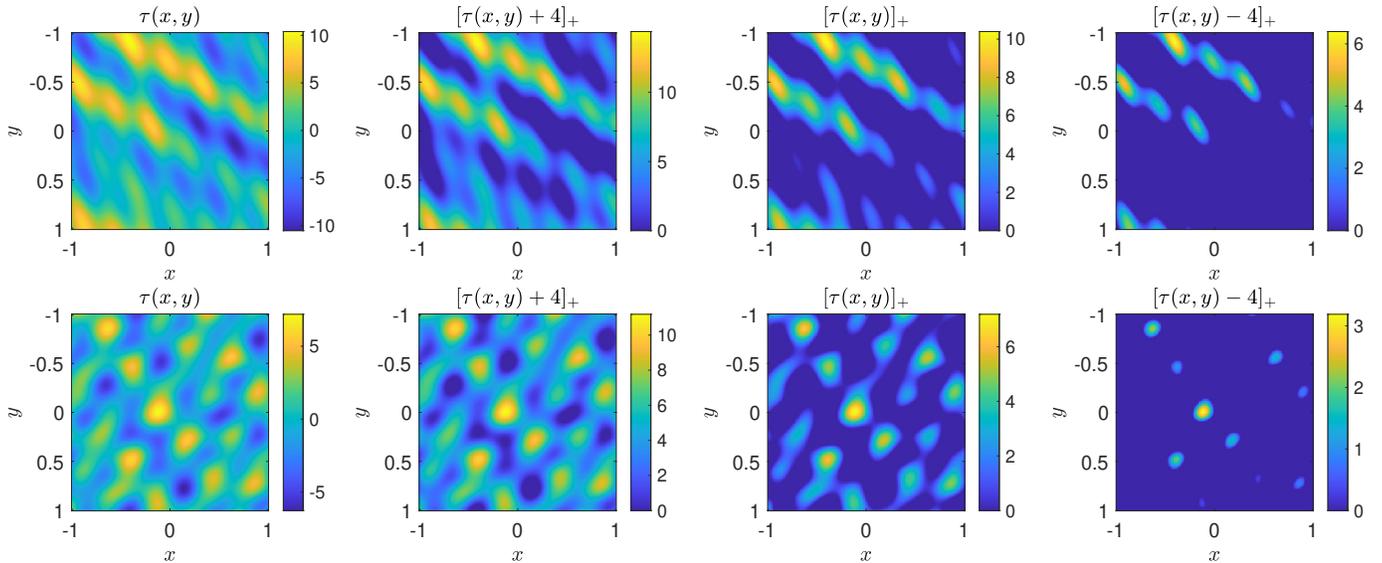}\\
    \caption{The single-hidden layer INR architecture considered in this work represents an image $f$ as a linear combination of rectified trigonometric polynomials $[\tau_i]_+ = [\vw_i^\T \bm\gamma(\vx)]_+$. We study the sample complexity of estimating an image of this type from its low-pass Fourier coefficients.}
    \label{fig:RTP}
\end{figure}
\subsection{INR Architecture}

We focus on an INR architecture first proposed in \cite{tancik2020fourier} that combines a trainable fully connected network with a Fourier features layer. For simplicity, we primarily consider a ``shallow'' INR architecture, where the fully connected network has a single-hidden layer with ReLU activation.

Specifically, consider a single-hidden layer ReLU network $g_\theta:\R^D\rightarrow \R$ of width-$W$ as given by
\begin{equation}\label{eq:MLP}
g_\theta(\vv) = \sum_{i=1}^W a_i[\vw_i^\T \vv]_+,
\end{equation}
where $[\cdot]_+$ denotes the ReLU activation function, and $\theta = \{(a_i,\vw_i)\}_{i=1}^W$ denotes the vector of all trainable parameters, which belongs to the space $\Theta_W \simeq \R^{(1+d)DW}$.
By choosing $D = d$, this architecture could be used as an INR directly. However, as shown in \cite{tancik2020fourier}, it is challenging to recover high-frequency content in images using such an architecture.

To enhance recovery of high-frequency content, \cite{tancik2020fourier} proposed adding an initial Fourier features layer to the architecture. 
In this work, we focus on a Fourier features layer $\bm\gamma:\R^d\rightarrow \R^D$ of the form
\begin{multline}
\bm\gamma(\vx) = 
[1, 
\sqrt{2}\cos(2\pi\vk_1^\T\vx), 
\cdots,
\sqrt{2}\cos(2\pi\vk_p^\T\vx),\\
\sqrt{2}\sin(2\pi\vk_1^\T\vx),
\cdots, 
\sqrt{2}\sin(2\pi\vk_p^\T\vx)
]^\T,
\end{multline}
where $\Omega_0 = \{\vk_1,...,\vk_p\}\subset \mathbb{R}^d$ is a collection of frequency vectors.  For a single hidden-layer fully connected network, this gives the INR architecture:
\begin{equation}\label{eq:ftheta}
f_\theta(\vx) = g_\theta(\bm\gamma(\vx)) = \sum_{i=1}^W a_i[\vw_i^\T \bm\gamma(\vx)]_+.
\end{equation}
To better understand the impact of the Fourier features layer,
for any weight vector $\vw \in \R^{D}$, let us write $\vw = [w^{(0)}, \vw^{(1)}, \vw^{(2)}]$  where  $w^{(0)} \in \R$ and $\vw^{(1)}, \vw^{(2)} \in \R^p$. Then defining $\tau(\vx) := \vw^\T \bm\gamma(\vx)$, we have
\[
\tau(\vx) = w^{(0)} + \sqrt{2}\sum_{j=1}^{p} \left(w_{j}^{(1)} \cos(2\pi\bm\vk_{j}^\T\vx) + w_{j}^{(2)} \sin(2\pi\bm\vk_j^\T\vx)\right).
\]
Note that $\tau$ is a \emph{trigonometric polynomial}, i.e., $\tau$ is a bandlimited function with frequency support contained in the discrete set $\Omega_0$.
Therefore, a single-hidden layer INR with Fourier features layer is a weighted linear combination of functions of the form $\vx \mapsto [\tau(\vx)]_+$. We call such a function a \emph{rectified trigonometric polynomial}; see Figure \ref{fig:RTP} for an illustration.

Originally, \cite{tancik2020fourier} proposed defining the frequency set $\Omega_0$ by sampling normally distributed random vectors $\vk \sim \gN(0,\sigma^2\mI)$, where the variance $\sigma$ is treated as a tunable parameter. To simplify the mathematical theory, in this work, we focus on the case where $\Omega_0$ consists of uniform sampled integer frequencies with maximum frequency $K_0$, i.e., $\Omega_0 =\{\vk \in \mathbb{Z}^d: \|\vk\|_{\infty}\leq K_0\}$. This implies that the INR is a periodic function on $[0,1]^d$. Using integer frequencies in place of random frequencies for an INR Fourier features layer was also investigated in \cite{benbarka2022seeing}.

\subsection{Generalized Weight Decay Regularization}\label{sec:weight_decay}

We focus on a family of parameter space regularizers $R(\theta)$ that strictly generalize ``weight decay'' regularization \cite{krogh1991simple}, i.e., the squared $\ell^2$-norm of all trainable parameters. Consider any non-negative weighting function $\eta:\R^D\rightarrow \R$ defined over inner-layer weight vectors $\vw \in \R^D$. We say $\eta$ is an \emph{admissible} weighting function if it is continuous, positively homogeneous (i.e., $\eta(\alpha \vw) = \alpha \eta(\vw)$ for all $\alpha \geq 0$), and $\eta(\vw) > 0$ whenever $[\vw^\T\bm\gamma(\cdot)]_+ \neq 0$. Given any admissible weight function, for any parameter vector $\theta = \{(a_i,\vw_i)\}_{i=1}^W$ we define the regularizer $R(\theta)$ by
\begin{equation}\label{eq:param_space_reg}
    R(\theta) = \frac{1}{2}\sum_{i=1}^W\left(|a_i|^2 + \eta(\vw_i)^2 \right).
\end{equation}
For example, if $\eta(\vw) = \|\vw\|_2$ we recover standard weight decay regularization. For the sampling theory given in Section \ref{sec:mainthm}, we focus on the alternative weighting function
\begin{equation}\label{eq:mwdreg}
    \eta(\vw) = \|\gF_\Omega[\vw^\T\bm\gamma(\cdot)]_+\|_2,
\end{equation}
and call the resulting $R(\theta)$ \emph{modified weight decay regularization}. Assuming $\Omega$ is a sufficiently large set of frequencies, by Parseval's Theorem we have $\|\gF_\Omega[\vw^\T\bm\gamma(\cdot)]_+\|_2 \approx \|[\vw^\T\bm\gamma(\cdot)]_+\|_{L^2([0,1]^d)}$. This shows that modified weight decay regularization can be seen as approximately penalizing the $L^2$-norm contribution made by each unit $[\vw^\T\bm\gamma(\cdot)]_+$ to the overall image.

\section{Theory}

Our main theoretical insight is to show that minimizers of the non-convex INR training problem \eqref{eq:opt_param_space} coincide with minimizers of a \emph{convex} optimization problem defined over an infinite-dimensional space of measures. We use this equivalence to prove our main theorem, which identifies a sufficient number of Fourier samples needed to recover a function realizable as a width-1 INR (i.e., a truncated trigonometric polynomial). Proofs of all results in this section are omitted for space reasons.

\subsection{Generalized Weight Decay Induces Unit Sparsity}
Let $\Theta^*_{W} \subset \Theta_W$ denote the subset of parameter vectors $\theta = \{(a_i,\vw_i)\}_{i=1}^W$ that satisfy the normalization constraint $\|\vw\|_2 = 1$ for all $i\in[W]$. Then we have the following equivalence:
\begin{myLem}\label{lem:ell2_to_ell1}
The function space minimizers of \eqref{eq:opt_param_space} coincide with the function space minimizers of
\begin{equation}\label{eq:opt_finite_width_R}
\min_{\substack{\theta \in \Theta_{W}^*}} \sum_{i=1}^W |a_i|\eta(\vw)~s.t.~\gF_\Omega f_\theta = \vy.
\end{equation}
\end{myLem}
The proof follows by an argument similar to the proof of \cite[Lemma 1]{savarese2019infinite}, and relies on the positive homogeneity of the ReLU activation and of the weighting function $\eta$. This result shows that by minimizing weight decay we implicitly minimize a weighted $\ell^1$-norm on the outer-layer weights $a_i$ of the network subject to a normalization constraint on the inner-layer weight vectors $\vw_i$.  Due to the sparsity-promoting property of the $\ell^1$-norm, intuitively this result shows that by minimizing $R(\theta)$ we should promote functions that are realizable by a ``sparse'' INR, i.e., one with few active ReLU units.

\subsection{Infinite-width Convexification}

One technical hurdle to characterizing minimizers of \eqref{eq:opt_finite_width_R} is that set of parameters $\theta \in \Theta_{W}^*$ satisfying the constraint $\gF_\Omega f_\theta = \vy$ is non-convex. To circumvent this issue, we show how to reformulate \eqref{eq:opt_finite_width_R} as a convex optimization problem posed over a space of measures that represents the limit as the hidden-layer width $W\rightarrow \infty$. We note that similar convex formulations for single hidden-layer neural networks have been explored in \cite{bengio2005convex,bach2017breaking,ergen2021convex}.

First, observe that for any $\theta = (\{a_i,\vw_i)\}_{i=1}^W\in \Theta_{W}^*$, by linearity we have
\[
\gF_\Omega f_\theta = \sum_{i=1}^W a_i \gF_\Omega[\vw_i^\T\bm\gamma(\cdot)]_+.
\]
We now show the sum on the right can be expressed as the integration of a function against a signed measure. In particular, define $\gU_\eta = \{\vw \in \R^D : \|\vw\|_2 =1, \eta(\vw) > 0\}$. For any finite signed measure $\mu$ defined over $\gU_\eta$, define the linear operator
\[
\gK_\Omega \mu := \int_{{\gU}_\eta} \gF_\Omega[\vw^\T\bm\gamma(\cdot)]_+ d\mu(\vw)
\]
Then for the sparse measure $\mu^* = \sum_{i=1}^W a_i \delta_{\vw_i}$ where $\delta_\vw$ denotes a Dirac delta measure centered at $\vw$, we have $\gF_\Omega f_\theta = \gK_\Omega \mu^*$ and so the constraint $\gF_\Omega f_\theta = \vy$ in \eqref{eq:opt_finite_width_R} is equivalent to $\gK_\Omega \mu^* = \vy$.

Next, define the weighted total variation norm  $\|\mu\|_{TV,\eta}$ for any signed measure $\mu$ over ${\gU}_\eta$ by
\[
\|\mu\|_{TV,\eta} = \int_{\gU_\eta} \eta(\vw) d|\mu|(\vw),
\]
where $|\mu|$ is the total variation measure associated with $\mu$. 
Then we have
$\|\mu^*\|_{TV,\eta} = \sum_{i=1}^K |a_i|\eta(\vw_i)$
which coincides with the objective in \eqref{eq:opt_finite_width_R}. 

Therefore, letting $\gM_W({\gU}_\eta)$ denote the set of measures defined over ${\gU}_\eta$ expressible as a weighted linear combination of at most $W$ Diracs, we have shown that \eqref{eq:opt_finite_width_R} is equivalent to the optimization problem
\begin{equation}\label{eq:opt_discrete_measures}
P^*_W = \min_{\mu \in \gM_W({\gU}_\eta)} \|\mu\|_{TV,\eta}~~s.t.~~\gK_\Omega \mu = \vy.
\end{equation}
In particular, if the measure $\mu^* = \sum_{i=1}^W a_i \delta_{\vw_i}$ is a minimizer of \eqref{eq:opt_discrete_measures}, then the parameter vector $\theta^* = \{(a_i,\vw_i)\}_{i=1}^W \in \Theta_{W}^*$ is a minimizer of \eqref{eq:opt_finite_width_R}, and hence by \Cref{lem:ell2_to_ell1}, the INR $f_{\theta^*}$ is also a function space minimizer of the original INR training problem \eqref{eq:opt_param_space}. This shows that we can describe all function space minimizers \eqref{eq:opt_param_space} by identifying the minimizers of \eqref{eq:opt_discrete_measures}.

However, analyzing \eqref{eq:opt_discrete_measures} directly is challenging since domain $\gM_W(\gU_\eta)$ is not a vector space.
To circumvent this issue, we pass to the larger space $\gM(\gU_\eta)$ of signed measures defined over $\gU_\eta$ with finite total variation norm, and consider the optimization problem
\begin{equation}\label{eq:opt_func_space}
    P^* = \min_{\mu \in \gM(\gU_\eta)} \|\mu\|_{TV,\eta}~~s.t.~~\gK_\Omega \mu = \vy.
\end{equation}
This can be thought of as the ``infinite-width'' analogue of \eqref{eq:opt_finite_width_R}. Note  \eqref{eq:opt_func_space} is now a convex optimization problem, though one posed over an infinite dimensional space of measures. 

Since $\gM_W({\gU}_\eta) \subset \gM({\gU}_\eta)$ we have $P^* \leq P_W^*$. Additionally, if a minimizer $\mu^*$ to \eqref{eq:opt_func_space} is a finite linear combination of at most $W$ Diracs, i.e., $\mu^* \in \gM_W(\gU_\eta)$, then $\mu^*$ must also be a minimizer of  \eqref{eq:opt_discrete_measures}, and so $P^* = P_W^*$. This is the key property we exploit in proving the sampling guarantees given below.

\subsection{Provable Recovery of Width-1 INRs}\label{sec:mainthm}
Now we give our main theorem, which identifies a sufficient number of samples for which an image realizable as a width-1 INR is exactly recoverable as the unique global minimizer of \eqref{eq:opt_param_space}. For technical reasons, here we focus on the modified weight decay regularizer $R(\theta)$ defined using the specific weighting function $\eta(\vw)$ given in \eqref{eq:mwdreg}. Also, we assume the Fourier feature frequency set $\Omega_0 = \{\vk\in \mathbb{Z}^d: \|\vk\|_\infty \leq K_0 \}$, and for any positive integer $n$, we let $n\Omega_0$ denote the dilated frequency set $\{\vk \in \mathbb{Z}^d: \|\vk\|_\infty \leq nK_0\}$. Then we have the following result:

\begin{myThm}\label{thm:main}
Let $f(\vx) = a[\tau]_+$ where $\tau$ is a real-valued trigonometric polynomial with frequency support in $\Omega_0$. Suppose $\Omega\supseteq 3\Omega_0$, and $\vy = \gF_\Omega f$. Then $f$ is the unique function space minimizer of the INR training problem \eqref{eq:opt_param_space} using modified weight decay regularization.
\end{myThm}

The main idea of the proof is to pass to the convex dual of \eqref{eq:opt_func_space}, which reduces to a semi-infinite program, i.e., a finite dimensional linear program with infinitely many constraints. Similar to previous works that have focused on super-resolution of sparse spikes \cite{candes2014towards,poon2019multidimensional,eftekhari2021stable}, we prove unique recovery is equivalent to the construction of a particular dual optimal solution known as a \emph{dual certificate}. In the case of a width-1 INR, we explicitly construct the dual certificate.

More generally, we conjecture that Theorem \ref{thm:main} can be extended to the case of standard weight decay regularization (i.e., weighting function $\eta(\vw)=\|\vw\|_2$). This is verified empirically in the next section. Additionally, when $f$ is a width $W$ INR, we conjecture that sampling over the frequency set $3W\Omega_0$ is sufficient to ensure $f$ is the unique minimizer of \eqref{eq:opt_func_space}, both with standard and modified weight decay regularization.

\section{Experiments}

\subsection{Exact Recovery Experiments}

 To validate the proved and conjectured sampling guarantees, we generate random continuous domain images realizable as a single hidden-layer ``teacher network'' INR of varying widths $W$; Figure \ref{fig:RTP} shows an example for $W=3$. We investigate the maximum sampling frequency $K$ needed to recover the image by solving the optimization problem \eqref{eq:opt_param_space} using a ``student network'' INR with the same architecture but larger width. As an attempt to solve the equality constrained problem \eqref{eq:opt_param_space} we use the Augmented Lagrangian (AL) method (see, e.g., \cite[Ch 7]{nocedal1999numerical}). This results in a sequence of ``outer loop'' INR training problems, each of which approximately optimized with $5,000$ iterations of the Adam optimizer with a learning rate of $0.001$, for a maximum of 60 outer loop iterations.

For both the teacher and student networks, we use a Fourier features layer with maximum frequency $K_0 = 2$. We sweep over $K$ values from $2$ to $30$ in increments of $2$, while varying the teacher network width, $W$, from $1$ to $5$. For each regularization approach and each $(K,W)$ pair, we run $10$ random trials. The width of student networks is $100$. 
We say the recovery is ``exact'' if the image domain mean-squared-error (MSE) is less than $1\times 10^{-9}$ at any point during training.

\begin{figure}[ht!]
    \centering
    \includegraphics[width=\columnwidth,height=4cm]{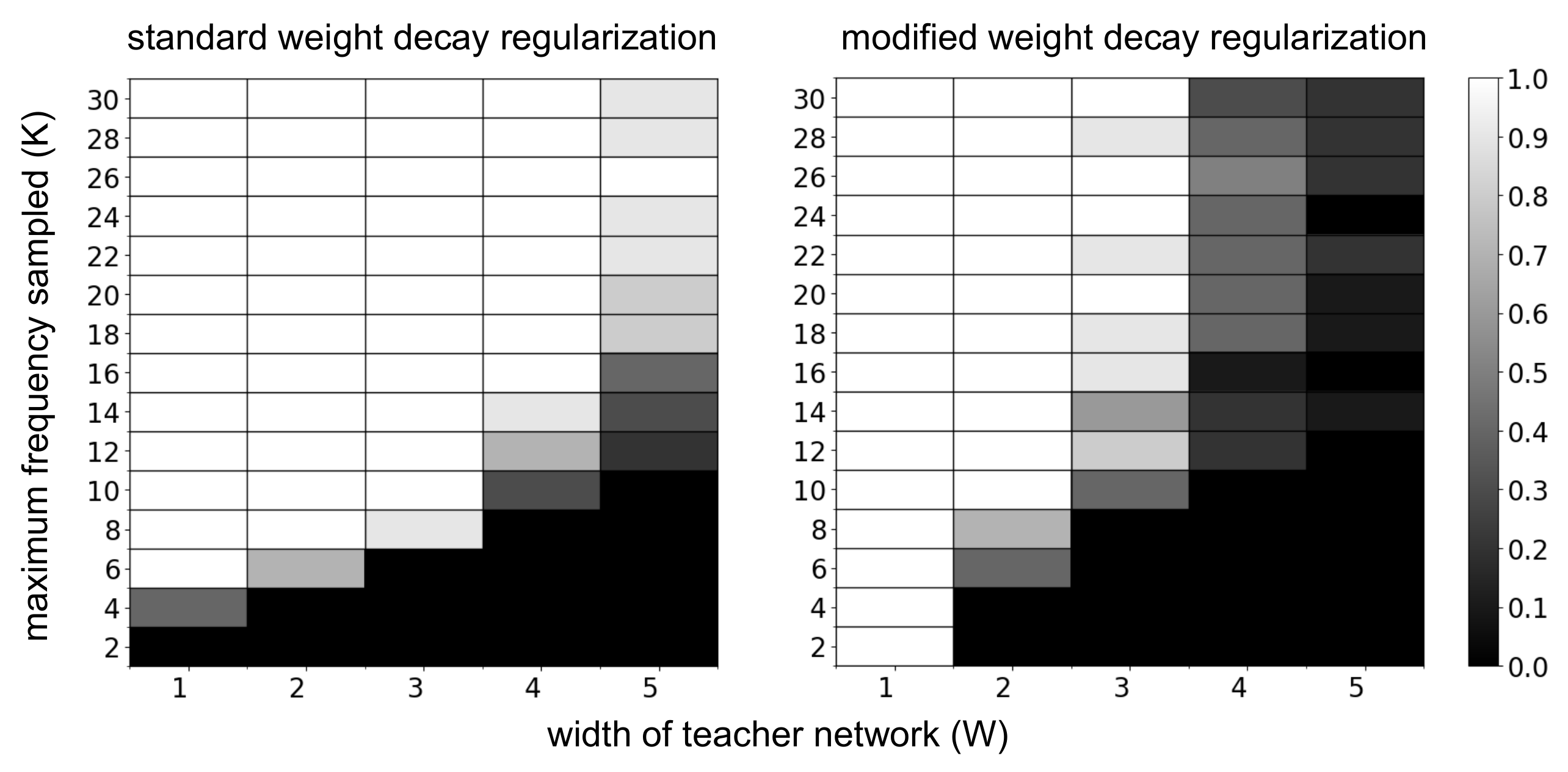}\\
    \caption{Empirical probability of exact recovery of images representable by a single hidden-layer INR of width-$W$ from its low-pass Fourier coefficients with maximum sampling frequency $K$ by solving \eqref{eq:opt_param_space} using standard and modified weight decay regularization.}
    \label{fig:exact_rec}
\end{figure}

\begin{figure*}[ht!]
    \centering
    \includegraphics[width=\textwidth]{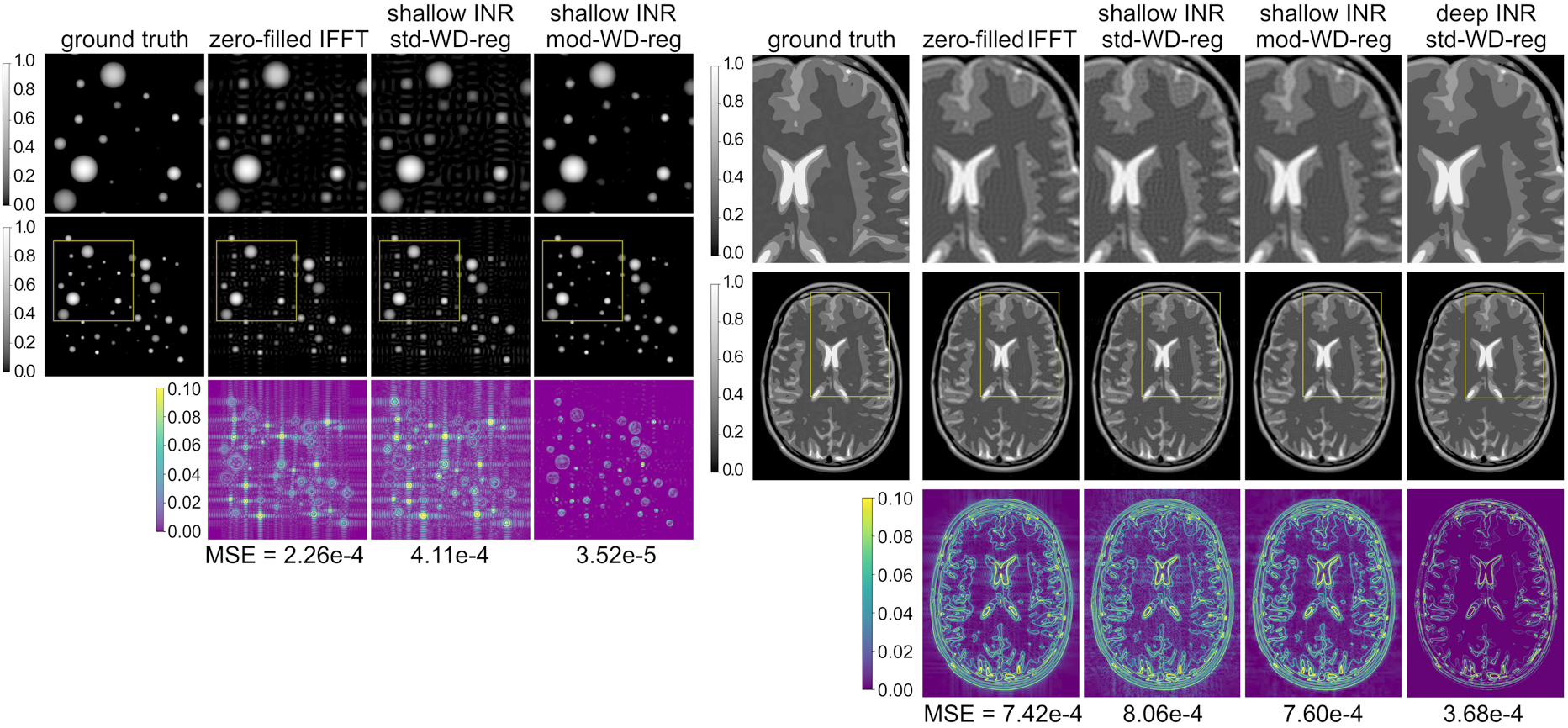}\\
        \vspace{-1em}
    \caption{\textbf{Recovery of continuous domain phantoms from low-pass Fourier samples.} Left panel:  ``dot phantom'' image created for this study features randomly generated circle-shaped features. Right panel: piecewise constant brain MRI phantom introduced in \cite{guerquin2011realistic}. We compare the zero-filled IFFT reconstruction with a single-hidden layer INR trained with standard weight decay regularization (shallow INR, std-WD-reg), and with the proposed modified weight decay regularization (shallow INR, mod-WD-reg). For the brain phantom, we also compare with a depth $10$ INR architecture trained with standard weight decay regularization (deep INR, std-WD-reg). The bottom row shows the absolute value of the difference with the ground truth.}
    \vspace{-0.5em}
    \label{fig:dot_brain}
\end{figure*}

Figure \ref{fig:exact_rec} shows the results of these experiments as probability tables. In the table corresponding to standard weight decay regularization, we see that the conjectured linear trend between the teacher network width $W$ and the minimal sampling cutoff $K$ needed to achieve exact recovery holds with high probability. For modified weight decay regularization, although the linear trend is less obvious, exact recovery is achieved in all random trials when $W=1$ and remains attainable in many cases when $W>1$. For both regularization methods, the parameter space optimization problem is non-convex, and so convergence of the AL method to a global optimum is not guaranteed. Therefore, it is unclear if the relatively worse empirical performance of modified weight decay regularization at achieving exact recovery is due to a difficulty in reaching a global minimizer with the proposed optimization method, or a more fundamental lack of identifiability. Despite this deficiency, we note there may be practical benefits to using modified weight decay over standard weight decay regularization in achieving \emph{approximate} recovery, as we show in the next set of experiments with more realistic phantom images.

\subsection{Recovery of Continuous Domain Phantom Images}
\vspace{-0.1em}
In Figure \ref{fig:dot_brain}, we illustrate the use of INRs for the super-resolution recovery of two continuous domain phantoms whose Fourier coefficients are known exactly: a ``dot phantom'' image created for this study that features randomly generated circle-shaped features, and a piecewise constant MRI ``brain phantom'' introduced in \cite{guerquin2011realistic}. 
The dot phantom is continuous, piecewise smooth, and exactly realizable by a shallow INR architecture of width $W \geq 50$ and $K_0 \geq 8$. Since the brain phantom is piecewise constant, and hence discontinuous, it is not exactly realizable with the INR considered in this work.
We also note both phantoms are not bandlimited, such that a simple reconstruction obtained by zero-filling in the Fourier domain and applying the IFFT results in significant ringing artifacts (see the second column in each panel of Figure \ref{fig:dot_brain}).

For the dot phantom, the maximum sampling frequency is set to $K=32$, and we fit a shallow INR with maximum frequency $K_0=10$ for the Fourier features layer and width $100$. For the brain phantom, the maximum sampling frequency is set to $K=64$, and we fit a shallow INR of width $100$ with maximum frequency $K_0=10$ for the Fourier features layer. To illustrate the role of INR depth, in the case of the brain phantom, we also fit a deep INR with depth $10$ and width $100$ in each layer. We train the INR by minimizing the regularized least squares formulation \eqref{eq:lsfit} using the Adam optimizer for $40,000$ iterations with a learning rate of $0.001$ followed by an additional $10,000$ iterations with a learning rate of $0.0001$. In each case, the regularization parameter $\lambda$ was selected via grid search to minimize image domain MSE. 

Unlike the exact recovery experiments, for both phantoms, we find that using a shallow INR trained with modified weight decay regularization results in lower image MSE and fewer artifacts in the reconstructions compared to using a shallow INR trained with standard weight decay regularization. In particular, the MSE is a magnitude lower on the dot phantom with modified weight decay and the reconstruction has far fewer ringing artifacts. A more modest improvement is shown on the brain phantom, where training with modified weight decay yields a slight improvement in MSE, but some ringing artifacts are still visible. However, we find that using a deep INR leads to further improvements in the brain phantom's recovery, eliminating nearly all residual ringing artifacts.

\section{Conclusion}
In this work, we introduce a mathematical framework for studying exact recovery of continuous domain images from low-pass Fourier coefficients with INRs. We show that fitting a single-hidden layer INR architecture using a generalized weight decay regularization term is equivalent to minimizing a convex loss over a space of measures. Using this perspective, we identify a sufficient number of samples to uniquely recover an image realizable as a width-1 INR, which we conjecture generalizes larger widths, as well.

While this work focused on a specific INR architecture introduced in \cite{tancik2020fourier}, several others have been proposed \cite{sitzmann2020implicit, muller2022instant, saragadam2023wire,shenouda2024relus}. Extending the present results to these alternative architectures is an interesting direction for future work.
Also, for mathematical tractability, we focused on a single hidden-layer INR architecture, but many studies use INR architectures with two or more hidden layers. Extensions of the theory to deeper architectures will also be a focus of future work.

\newpage
\section*{Acknowledgment}
This work was supported by NSF CRII award CCF-2153371.

\bibliographystyle{IEEEtran}

\bibliography{IEEEabrv,references}

\end{document}